\RequirePackage{fix-cm}
\documentclass[smallextended]{svjour3}       %
\smartqed  %
\usepackage{graphicx}
\usepackage[utf8]{inputenc}
\usepackage[english]{babel}
\usepackage[labelfont=bf,labelsep=quad]{caption} %
\usepackage{amsmath,amssymb}
\usepackage{tabularx}
\usepackage{array}
\usepackage{booktabs}
\usepackage{threeparttable}
\usepackage{setspace}
\usepackage{listings}
\usepackage{xcolor}
\usepackage{multirow}
\newcolumntype{P}[1]{>{\raggedright\arraybackslash}p{#1}}
\usepackage{graphicx}
\usepackage{float}
\usepackage{enumitem}
\usepackage{wrapfig}
\PassOptionsToPackage{hyphens}{url}\usepackage[colorlinks=true,pdfborder={0 0 0},linkcolor=blue,filecolor=blue,citecolor=blue,urlcolor=blue]{hyperref}
\definecolor{UniBlue}{RGB}{0, 74, 153}
\definecolor{RandomRed}{RGB}{181, 93, 76}
\usepackage[none]{hyphenat}
\usepackage[]{natbib}

\usepackage{rotating}

\newcolumntype{L}[1]{>{\raggedright\arraybackslash}p{#1}}
\newcolumntype{C}[1]{>{\centering\arraybackslash}p{#1}}
\newcolumntype{R}[1]{>{\raggedleft\arraybackslash}p{#1}}

\begin{document}

\title{Biases in Scholarly Recommender Systems: Impact, Prevalence, and Mitigation}
\author{
Michael Färber \and Melissa Coutinho \and Shuzhou Yuan
}
\institute{Michael Färber \at
           Institute AIFB, Karlsruhe Institute of Technology (KIT), Germany\\
           \email{michael.faerber@kit.edu}
           \and
           Melissa Coutinho \at
           Institute AIFB, Karlsruhe Institute of Technology (KIT), Germany\\
           \email{melcouts97@gmail.com}
           \and 
           Shuzhou Yuan \at
           Institute AIFB, Karlsruhe Institute of Technology (KIT), Germany\\
           \email{shuzhou.yuan@kit.edu}
           }
\date{Received: date / Accepted: date}
\maketitle

\begin{abstract}
With the remarkable increase in the number of scientific entities such as publications, researchers, and scientific topics, and the associated information overload in science, academic recommender systems have become increasingly important for millions of researchers and science enthusiasts. However, it is often overlooked that these systems are subject to various biases. In this article, we first break down the biases of academic recommender systems and characterize them according to their impact and prevalence. In doing so, we distinguish between biases originally caused by humans and biases induced by the recommender system. Second, we provide an overview of methods that have been used to mitigate these biases in the scholarly domain. Based on this, third, we present a framework that can be used by researchers and developers to mitigate biases in scholarly recommender systems and to evaluate recommender systems fairly. Finally, we discuss open challenges and possible research directions related to scholarly biases. 
\end{abstract}

\section{Introduction}
\label{Label1}

{\textit{Bias}  
refers to the phenomenon of unfairly favoring a group of people or an opinion \citep{o2016weapons, delgado2004bias,9706439}, %
and is 
highly relevant for
recommender systems. 
For instance, in the case of movie recommendation, 
users tend to rate popular 
movies more often, which results in unpopular
movies being recommended less frequently \citep{Park2008TheIt,Abdollahpouri2017ControllingRecommendation}. 
\citet{Liu2016AreModeling} found that conformity, which implicitly shapes people’s behaviors to group
norms, strongly influences users' rating behavior in 
recommender systems. 
\citet{Zehlike2017FAIR:Algorithm} found that people search engines, quite commonly used in job recruiting and to find friends on social media, are biased based on gender, race, and physical disabilities. Identifying and mitigating bias in AI systems, particularly recommender systems, has therefore been garnering interest from researchers worldwide. For instance, some have proposed new learning-to-rank algorithms to mitigate bias \citep{Zehlike2017FAIR:Algorithm,Abdollahpouri2019ManagingRe-Ranking}, while others have proposed techniques to better understand and model user preferences for recommendation \citep{Zheng2020DisentanglingEmbedding,Liang2016}.

In the scholarly domain, researchers are faced with excessive information overload (e.g., tens of thousands of papers published annually in specific fields \citep{farber2019microsoft}). 
Consequently, scholarly recommender systems---which recommend papers \citep{Bogers2008RecommendingCiteULike}, research trends \citep{Zhao2018AcademicSharing}, collaborators \citep{Salman2020IncorporatingRecommendation}, journals, conferences \citep{Klamma2009}, and other entities---are increasingly gaining attention. However, to the best of our knowledge, scholarly bias has not been considered 
systematically 
so far. This is surprising, as 
academic society 
has an ethical responsibility \citep{Polonioli2020TheSystems}.
Scholarly bias can affect millions of researchers in various fields. It indirectly influences research output and funding decisions regarding research projects with cumulative multibillion dollar budgets. 
Bias affects which research ideas get promoted and are finally used in industries, impacting the long-term welfare of industrial nations. %
With the increase in the number of papers on bias in recommender systems in general, it is 
important that scholars 
have a global picture of the various types of 
bias 
in recommender systems and bias mitigation methods. 
In addition, current works on bias are fragmented and use important terms in different ways. For instance, some papers refer to ``selection bias'' as bias that occurs due to undersampling training data \citep{Wang2016,Zhang2019SelectionDatasets}, while others use that term 
to refer to the bias caused by limited user exposure to recommendation results \citep{Ovaisi2020CorrectingSystems}, which means only a small number of items are displayed to a user. Also, some papers do not mention the term ``bias'' in their content, but 
address at least one type of bias \citep{Liang2016,Sugiyama2013ExploitingRecommendation,Gai2014DualFeedbacks}. 
Due to these aspects, it might be difficult for researchers interested in this topic to find significant related work. In addition to this, it is difficult to identify and measure bias in scholarly recommender systems despite the pressing need to do so, because these systems face issues related to domain specificity, data sparsity, and data heterogeneity (see Sec.~\ref{sec:Label7}). 
Thus, we believe it is 
necessary to provide an overview of this area, so as to help researchers understand the current state 
and future work on this topic.  

\begin{figure} [tb]
    \centering
    \includegraphics[scale = 0.65]{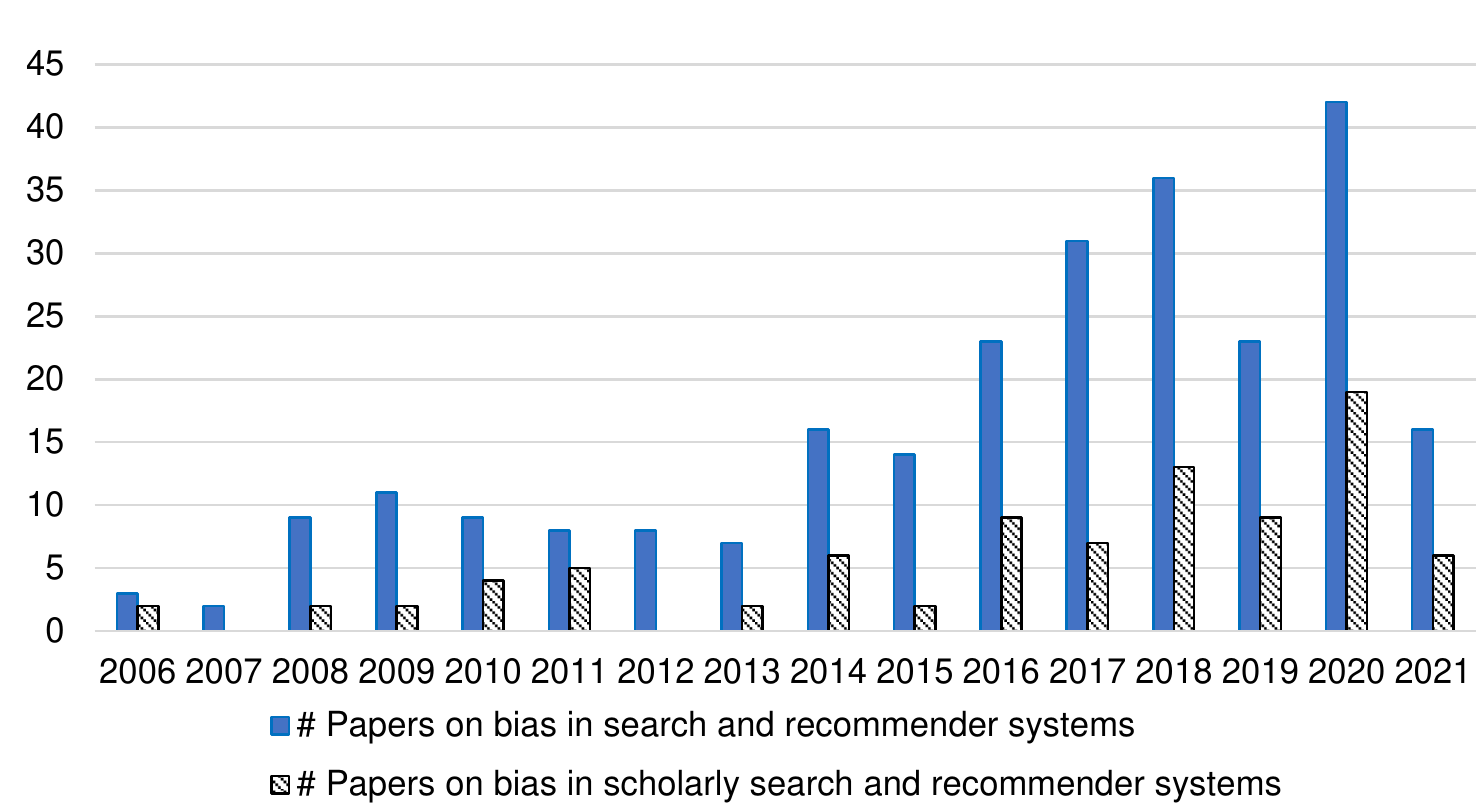} %
    \caption{Number of papers 
    in our corpus by publication year.} %
    \label{fig:bias_publication} 
\end{figure}

Biases in \textit{scholarly} recommender systems have been disclosed
in several works: 
For instance, \cite{Liang2016} discussed the recommender systems' ``exposure problem'', which can result in frequent recommendation of popular scientific articles. 
\cite{Salman2020IncorporatingRecommendation} observed gender and racial biases as well as biases based on location in academic expert recommendations used in the hiring process, to find reviewers, or to assemble a conference program committee. 
In addition, researchers speak of ``filter bubbles'' (``information bubbles'') as 
the phenomenon that people are isolated 
from a diversity of viewpoints or content 
due to online personalization \citep{Nguyen2014ExploringDiversity}.
\cite{Polonioli2020TheSystems} 
claimed 
that recommender systems might isolate users in information bubbles by insulating them from exposure to different academic viewpoints, creating a self-reinforcing bias damaging to scientific progress. Finally, \citet{Gupta2021CorrectingRecommendation} found that scholarly recommender systems are biased as they underexpose users to equally relevant items.  Figure~\ref{fig:bias_publication}\footnote{Statistics calculated based on the set of papers collected for our literature survey. See Section~\ref{sec:biases-rec} for more information.} shows the number of papers 
on bias in scholarly recommender systems and in recommender systems in general. 
We observe that bias both in recommender systems in general and in scholarly recommender systems has received a lot of attention. 
A 
paper 
that tackles the popularity bias of recommending scientific articles \citep{Wang2011CollaborativeArticles}  won the Test of Time award at the KDD 2021 conference.\footnote{\url{https://kdd.org/awards/view/2021-sigkdd-test-of-time-award-winners}} In addition, the best paper award at SIGIR in 2020\footnote{\url{https://sigir.org/sigir2020/awards/}} was awarded to  \cite{Morik2020ControllingLearning-to-Rank} for their paper on fairness in learning-to-rank systems. %
All this indicates that bias in scholarly recommender systems is a timely and important topic.

Few literature surveys in the context of bias in recommender systems \textit{in general} (i.e., independent of the domain) 
have been published \citep{Chen2020,Mehrabi2019ALearning,Piramuthu2012InputSystems}. 
For instance, \citet{Chen2020}'s survey 
highlighted 
the different biases a recommender system faces (excluding fairness), as well as various methods of mitigating them. Our survey differs from these works in the following aspects: 
\begin{enumerate}
 \item We focus on detecting and mitigating biases found in scholarly search and recommender systems. The scholarly domain has not been considered systematically with respect to biases, while it exhibits several peculiarities that make 
identifying and mitigating 
biases challenging and not straight forward. %
To name only a few peculiarities, which cannot be found in other domains, such as news publishing:
\begin{enumerate}[label=(\alph*)]
\item~Papers are domain-specific texts, using a specific vocabulary and often following a dedicated structure, such as IMRaD \citep{IMRAD}. Furthermore, papers
are enriched by citations. Both the prose and the citations are prone to bias. 
\item~Papers typically have several authors, often with different background (e.g., gender, institution, country), which can be a ground for bias.
\item~Research takes place in scientific communities, which are often quite separated from each other, leading to specific social norms. %
\item~Science thrives on scientific discourse between researchers. In addition, it is increasingly the case that papers are distributed and promoted on social media, which can lead to further effects and bias.
\item~Scholarly systems are known to suffer from cold-start problems \citep{Wang2011CollaborativeArticles}. The number of items is much greater than the number of users, leaving many items without any feedback. \cite{Vellino15} 
found 
that the sparsity of the user-item matrix in Mendeley was three times smaller than that of Netflix.
\item~The Matthew effect (``the rich gets richer'') was found to be particularly prevalent in academia \citep{Merton1968TheScience,10.1145/3308558.3313593,DBLP:journals/ecra/ZhouZWHW22,wang2021science} 
\end{enumerate}

\item 
We not only specify a definition of each bias, but also its impact and prevalence in the academic field, showing which biases are particularly relevant and worth to be investigated further. 
\item 
We propose
a framework on bias detection and mitigation 
in \textit{scholarly} recommender systems. %
\end{enumerate}

Overall, in this article, we make the following contributions:
\begin{enumerate}
    \item We define the types of biases 
    in scholarly 
    recommender systems, and 
    describe their characteristics and influence in academia. %
    \item We summarize existing methods for debiasing 
    scholarly recommender systems, and provide a framework for 
    applying these methods. 
    \item We identify open challenges and discuss future directions to inspire further research on scholarly bias. %
\end{enumerate}

The structure of this article is as follows: In Section \ref{sec:biases-rec}, we identify the various biases present in scholarly recommender systems  
and classify them according to their impact and prevalence. 
Section \ref{sec:methods-mitigate} explains methods of bias mitigation. 
It also proposes a debiasing framework that can be followed to detect and mitigate biases in scholarly recommender systems. Finally, Section~\ref{Label27} discusses future research directions and the challenges faced when building or using such systems.

\section{Recommendation Bias} %
\label{sec:biases-rec}

In this section, we first introduce definitions of important concepts used throughout the article.
We then describe 
how we obtained our collection of papers 
concerning scholarly biases, 
categorize types of scholarly bias according to the lifecycle of (scholarly) recommender systems, and describe the impact and prevalence of the individual biases.

\vspace{-0.3cm}
\textcolor{black}{\subsection{Terminology}}

In the following, we define important concepts used throughout our article.

\begin{itemize}
 \item \textcolor{black}{\textbf{Bias} refers to the phenomenon of unfairly favoring a group of people or an opinion \citep{o2016weapons}. \textbf{Scholarly bias} refers to \textit{bias} occurring in the scholarly domain (e.g., when dealing with scientific publications, researchers, affiliations, and venues), such as in scholarly recommender systems.} 

 \item \textcolor{black}{\textbf{Conformity} is an important factor leading to bias. It is defined as the act of matching attitudes, beliefs, and behaviors to group norms, politics or being like-minded \citep{cialdini2004social}.} 

 \item \textcolor{black}{\textbf{Filter bubble} or \textbf{information bubble} is a persistent phenomenon existing in search engine and social media \citep{bruns2019filter}. With the personalized recommendation, users are exposed to information targeting their interests and reinforcing their belief instead of seeing balanced information. Consequently, users are isolated in the bubbles, where only the information consistent with their viewpoints can be seen \citep{pariser2011filter}.}


 \item \textcolor{black}{\textbf{Echo chamber} is caused by the lack of diverse perspectives and framed by like-minded users, especially on social media platforms \citep{cinelli2021echo}.
 In an \textit{echo chamber}, a group of people having the similar perspective choose to preferentially connect with the people inside the group, and exclude the viewpoints of outsiders \citep{bruns2017echo}.}
\end{itemize}

\vspace{-0.2cm}
\subsection{Paper Collection}
\label{sec:corpus}

To collect relevant literature for this survey, we first performed a keyword search following \citet{Beel2016}'s procedure using Google Scholar as a data source.\footnote{Two separate searches were conducted in April and October 2021, respectively.} We relied on Google Scholar, as it is not only a commonly used academic search engine \citep{haddaway2015role}, 
but also covers more citations than other sources \citep{martin2021google}. We used ``bias,'' ``fairness,'' ``recommender'' or ``recommendation,'' combined with ``paper'' or ``citation,'' as keyword queries. 
Following the snowball sampling technique, we broadened the search by looking through the bibliography of these papers, and downloaded relevant papers 
that had been cited by 
the first set of works. 
To narrow down 
our results, we searched using keywords such as ``academic'' or ``scholarly'' in papers' titles and body text. 
Overall, 88 documents addressed some kind of bias in scholarly searches or recommender systems, while 171 documents addressed bias 
in general. 
The list of all considered publications is available online.\footnote{See \url{https://figshare.com/s/2b90f185d12067986ec2}.} 

\subsection{Categories of Bias in Scholarly Recommender Systems}
\label{Label5}

According to \cite{Chen2020}, we can differentiate between the following steps of recommender systems: 
\begin{enumerate}
    \item \textit{Training of recommender systems model (Data $\rightarrow$ Model)}: recommender systems are trained based on observed user-item interactions, user profiles, item attributes, etc.
    \item \textit{Recommending items to users (Model $\rightarrow$ User)}: recommender systems infer user preferences toward items and provide recommendations to the users. %
    \item \textit{Collecting data from users (User $\rightarrow$ Data)}: New user actions are integrated into the training data. 
\end{enumerate}

We argue that biases can be categorized more intuitively when considering their immediate cause -- which is either the human user or the recommender system 
itself. 
In the following, we  
outline the biases of these 
bias categories. 

\vspace{-0.35cm}
\subsection{Human-caused Bias} 
\label{Label6}

%
The human-caused bias 
category includes biases that occur outside of the recommender system's training and deployment steps and covers biases regarding the user behavior. 
User behavior can be collected 
explicitly via ratings \citep{Tang2009}
and implicitly through user-item interactions, such as clicks, downloads, and citations. While explicit data tells us explicitly whether a user likes or dislikes an item, it is often difficult and labor-intensive to obtain it \citep{Yang2009CARES:System,Naak2009APapyres}. Most recommender systems nowadays therefore use implicit feedback data \citep{Pennock2013,McNee2002}. If user decisions and actions are biased, this results in biased data as well. 
As recommender system models are typically trained on recorded user behavior (e.g., user interactions with an existing recommender system), 
the biases in data collected from users leads to biases in model training which could lead to the recommender systems model making biased predictions (recommendations) to users.


\textit{Conformation bias}, \textit{position bias}, and \textit{selection bias} are bias types  
which are mentioned by \cite{Chen2020} 
as biases that occur in recommender systems in general and  
which we consider as being immediately human caused. 
In the following, we outline what these biases mean in the scholarly domain in terms of their impact and prevalence. \\

\begin{enumerate}

   \item \textbf{Conformation bias}
    \label{Conformation}
    \par \textbf{Definition:} \textit{Users tend to conform their beliefs with those of their peers, friends, and those in the same communities, even if doing so goes against their own judgment, meaning the rating values do not always signify  true user preference \citep{Chen2020}.}
\par \textbf{Impact:} In case of the conformation bias, scholars ignore personal beliefs to follow those of their peers. Conformation bias can have a significant impact on the academic society, resulting in ethical and social concerns. \textcolor{blue}{It} can be understood as non-optimal scientific practice, as researchers should steadily rely on themselves and should take over responsibility for their actions. In particular, social media found its way into science, leading to the effect that scholars may prefer items favored by their social ties or communities. 
\par \textbf{Prevalence:} Researchers are exposed to a variety of social influences, including the urge to fit in with their peers. By evaluating a network epistemology model in which scientists tend to adopt activities that conform to those of their neighbors, \cite{Weatherall2020ConformityNetworks} investigated the interplay between conformism and the scientific community's epistemic goals. They came to the conclusion that conformism reduces the likelihood of actors taking successful activities %
\citep{Weatherall2020ConformityNetworks}.  
Studies have also found that the decision of a group can be biased toward the opinion of the group member who started the discussion.

Researchers use a variety of social media platforms \citep{van2018platform}, including ResearchGate, LinkedIn,\footnote{\url{https://www.linkedin.com/feed/}} Facebook,\footnote{\url{https://www.facebook.com/}} Twitter,\footnote{\url{https://twitter.com/}} and Academia,\footnote{\url{https://www.academia.edu/}} to communicate and disseminate information, such as sharing their own articles with possible readers and looking for collaborators. Scholars who have no paper access due to license restrictions, especially the ones in underdeveloped areas, may obtain paper resources in this way. According to a 2015 survey, 47\% of scientists affiliated with the American Association for the Advancement of Research (AAAS) use social media to keep up with new findings and discuss science by sharing their thoughts \citep{HowCenter}. The ability to create social media networks has aided in the communication and collaboration of scientists irrespective of geographical location \citep{2018SocialScientists}. 

Scholarly metrics like altmetrics\footnote{\url{https://www.altmetric.com/}} count the number of times a research paper has been downloaded or shared on social media networks like Twitter to determine its qualitative and quantitative impact \citep{Adie2013Altmetric:Metrics}. Some research also shows that the most cited scientific papers are also the ones with the highest impact according to altmetrics \citep{Torres-Salinas2013Altmetrics:2.0}. However, there might be remarkably exceptions in which altmetrics say nothing about the quality of artifacts. For instance, there are three retracted papers within the top 100 articles at altmetric.com as of November 2021.  Altmetrics indicators that use data from social networks are ever present in the research landscape. Due to the growing influence of social networks in our daily lives and to simultaneously tackle the data sparsity problem, recent studies tend to integrate social media information, such as altmetrics, into the recommender systems, known as social recommender systems \citep{Zhao2018AcademicSharing,Asabere2014ScholarlyFolksonomy,Xu2012CombiningRecommendation,Lee2011ImprovingSystem}.  \cite{Wang2017LearningRecommendation} offer a social recommendation model that indicates if individuals have affinities with items favored by their social ties. %
The conventional explanation is that a user's taste is comparable to and/or affected by her trustworthy social network friends. %
\\
Scholars 
may look for information for a quick conclusion, or cease seeking information once a conclusion is reached. In such cases, personalized social recommender systems that use social information to create recommendations could promote conformity in scholarly circles \citep{Polonioli2020TheSystems}. Furthermore, \cite{Analytis2020TheNetworks} found that traditional collaborative filtering algorithms for recommender systems, such as the weighted k-nearest neighbors (k-nn) algorithm, produce social influence networks, and that the most influential individuals (e.g., highly cited researchers or famous scientists) benefit the most from the k-nn algorithm. They examined datasets from Faces \citep{DeBruineLisaJones2017FaceSet} and Jester \citep{Goldberg2001Eigentaste:Algorithm} and show that as the number of users in a network increases, so does the influence of only a handful of people.  \cite{Zheng2020DisentanglingEmbedding} propose a framework to disentangle user's interest and conformity because a user might click an item, not because she likes it, but because many others have clicked on it. This is true for any platform that uses a social metric, such as likes, downloads, and citations. Therefore, influenced by other's opinions, users' rating values might not signify true preference. Explicit, published indicators or analyses 
on the prevalence of conformation bias in scholarly datasets were not found by us.\\

\newpage{}
\item \textbf{Position bias} 
\par \textbf{Definition:} \textit{Users tend to interact with items in higher position of the recommendation list regardless of the items’ actual relevance \citep{Collins2018ASystems}.}

\textbf{Impact:} 
Position bias makes the evaluation of recommender systems challenging, because it also affects how the users interact with a series of recommendations \citep{mansoury2021understanding}. Tracking a user's clicks on a series of recommendations can often be used to infer the relevance of a set of recommendations. This click data can thus be used to assess the effectiveness of a recommender system. However, due to position bias, the likelihood of a user engaging with an item may not reflect its absolute relevance within the set. 
If position bias is not corrected 
in scholarly search and recommender systems, some recommended items
(papers, venues, citations, etc.) will receive less than their ``fair'' share of clicks, views, downloads, etc. and others will receive more than their ``fair'' share. 
Considering the classical \textit{Matthew effect}\citep{Merton1968TheScience}, %
researchers also tend to use the papers that show up in the top $n$ for the purpose of citing. Subsequently, those papers receive more citations and are weighted higher in further search results.

\par \textbf{Prevalence:} In 
search and recommender systems (e.g., Google Scholar search), position bias seems to be a widespread phenomenon: 
According to eye-tracking studies \citep{Joachims2017AccuratelyFeedback}, users are less inclined to look at lower-ranking items in vertical lists, instead focusing on the first few entries. Furthermore, 65\,\% of users engage with lists in a depth-first manner, clicking on the first item that appears relevant without thoroughly evaluating the full list \citep{KlocknerDepth-andLists}. When \cite{Joachims2017AccuratelyFeedback} studied the user behavior on Google's results page (see Fig.~\ref{Position_bias}), they found that users tend to click substantially more often on the first rather than the second ranked item, while they view both items with almost equal frequency. This behavior is similar when using academic search engines like Google Scholar, PubMed, Web of Science\footnote{\url{https://www.webofknowledge.com/}}, etc. where the items (e.g., papers) ranked higher receive higher citations, views, clicks, downloads, etc. \cite{Wang2018PositionSearch} also observed a considerable position bias when users search for emails. \\

    \begin{figure}[tb]
    \centering
    \includegraphics[width=0.7\textwidth]{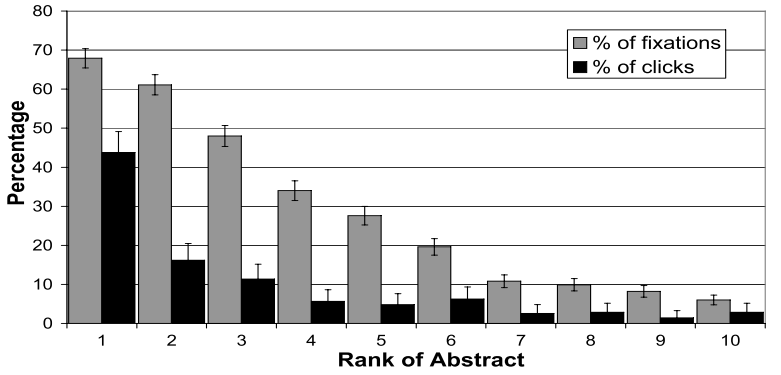} 
    \caption{Figure from \cite{Joachims2017AccuratelyFeedback}. Percentage of times an abstract was viewed (``fixations'') and clicked depending on the result rank of the result. The abstracts ranked 1 and 2 receive most attention.} %
    \label{Position_bias}
    \end{figure}
\item \textbf{Selection bias}
\par \textbf{Definition:} \textit{Users sometimes choose to rate items not out of relevance or item quality, but simply because they are influenced by other factors like item popularity, higher citation count of author, paper, etc. which have little or nothing to do with relevance or quality. As a result, observed ratings are not a representative sample of all ratings \citep{Chen2020}.}
\par \textbf{Impact:} Selection biases in the data impact search and recommender systems for scholars by making inaccurate predictions. In recommender systems the algorithms used to predict user preferences are designed to have high prediction accuracy on the assumption that the missing implicit ratings are missing at random (MAR), i.e., there is no bias operating over which items are rated and which are not. MAR signals rarely exist in the real world, because it is very unlikely that a recommender system would recommend items completely at random. Implicit data can be missing-not-at-random (MNAR) due to the exposure, position, and popularity bias of the recommender systems model \citep{Stinson2021AlgorithmsFiltering}. This leads to a \textit{skewed observed rating distribution}, because there is a lot of data on certain (popular, relatively old) items, whereas there is very less data on other (state of the art, relatively new) items. This makes learning the preferences of users challenging for a recommender system model. Furthermore, recommender system algorithms that depend entirely on observed implicit data or naively based on such missing click data will produce biased recommendations \citep{Hu2008CollaborativeDatasets,Steck2010TrainingRandom,Marlin2009CollaborativeData}. It is also completely up to the user which item she chooses to rate (in the form of clicks, views, downloads, etc.). She might choose to rate an item not because she finds it relevant, but because of item popularity, high citation count, etc. Since her selections are biased, the observed rating distributions are biased as well. For instance, most scholars would choose to cite a paper that has many citations, or one that was presented at a world-renowned conference, over another paper that has fewer citations, even though the content of the latter was more relevant to them. As another example, papers written by a famous author are more likely to be viewed regardless of relevance or interest, simply because most scholars tend to access them. \cite{wang2021science} revealed that in the academic world, famous scientists get credit easier than less famous researchers. This could lead to items being labeled as false positives in the training data and the recommender system making incorrect predictions. Selection bias also occurs in the \textit{sampling process}, when the chosen data sample is biased and not a representative of the whole population \citep{Zadrozny2004LearningBias}. 
\par \textbf{Prevalence:} Selection bias can be observed in IR systems of various kind, such as paper search engines (e.g., Google Scholar) and paper/citation/dataset recommender systems. 
Note that position bias (i.e., bias of considering only top-k results) can also lead to selection bias (i.e., bias due to considered background information) in such systems \citep{Ovaisi2020CorrectingSystems}. Users rarely 
view all relevant results, either because the system displays a shortened list of the top-k recommended items or because users do not take the time to review all of the ranked results. Lower ranked, relevant results have a marginal  
chance of being noticed (and clicked) 
and may never be boosted in Learning to Rank (LTR) systems, where a ranking model is learned based on training data. \cite{Ovaisi2020CorrectingSystems}  therefore model selection bias in two semi-synthetic datasets from the Yahoo! Learning to Rank Challenge (C14B)\footnote{\url{https://webscope.sandbox.yahoo.com/catalog.php?datatype=c}} by assigning a zero observation probability to documents below a cutoff $k$. In the scholarly domain, \cite{Forsati2017Semi-supervisedTop} investigated a selection bias in the CiteULike\footnote{\url{https://www.citeulike.org}} dataset, containing user ratings for papers. 

Selection bias is relatively comprehensively diagnosed 
in non-scholarly datasets: \cite{Mena-Maldonado2021PopularityEvaluation} observed this bias in ratings of the MovieLens 1M dataset \citep{Maxwell2015TheDatasets}, as the small number of popular items take up most of existing ratings, thereby displaying a MNAR pattern in the data distribution. Selection bias was also found in the Yahoo! base rating dataset \citep{Marlin2012CollaborativeAssumption}, where the users have a probability of rating either 1 (very good) or 5 (very bad) more than 50\% of the time. \cite{Zhang2019SelectionDatasets} observed selection bias in a Natural Language Sentence Matching (NLSM) dataset like the QuoraQP dataset during the sampling process. Quora\footnote{\url{https://www.quora.com/}} is a social platform where knowledge can be shared by scholars and non-scholars alike. The authors analyzed how the sampling procedure of selecting some pairs of sentences by the providers can bring about an unintended pattern, i.e. selection bias, into the model. We were not able to discover explicit statistics on the prevalence of selection bias in scholarly datasets. \\

\end{enumerate}
\subsection{System-caused Bias}
\label{sec:Label7}

\textit{System-caused bias}, which we can also name \textit{model-intrinsic bias}, occurs during the training and deployment of recommender systems. It is caused by unfair data or inducted by the model itself. If the data with bias is used for training a recommender systems model, the bias might also be in the recommender system's results. Due the effects of the feedback loop (i.e., recorded user behavior with the recommender system is used to retrain the recommendation model), the bias in the data might increase even more, leading to the ``rich get richer'' Matthew effect \citep{Chaney2015ARecommendation}. Moreover, unfair recommended results caused due to biased and imbalanced data might also have a collective, disparate impact on certain groups of people \citep{Courtland2018} like female scholars, scholars from developing countries, new papers published by relatively less experienced scholars. This \textit{unfairness} leads to gender and racial biases among many others, which can cause detrimental impact to scholars looking for jobs and promotions, grant funding (with a total budget of more than 2\% of the countries' GDP worldwide \citep{worldgdp}), etc.

In the following, we outline the system-caused biases in the context of scholarly recommender systems. Following mainly \cite{Chen2020}, who described bias in recommender systems in general, these are \textit{inductive bias}, \textit{exposure bias}, \textit{popularity bias}, and \textit{unfairness}.


\begin{enumerate}
    \item \textbf{Inductive bias (Learning bias)}
    
\par \textbf{Definition:} \textit{The model makes assumptions 
to better learn the target function and to generalize the training data \citep{Chen2020}}.
\par \textbf{Impact:}
Strictly speaking, inductive bias is not part of the traditional definition of bias as unfairly favoring a group of people or an opinion \citep{o2016weapons, delgado2004bias,9706439}. 
Inductive bias is considered to be positive, 
as it leads to more accurate results. 
It is added to the model to enable that the solution learned from the observed training data can better adjust to unseen data in the real world and not only to the specific test data. To achieve this, often a regularization term is introduced to avoid overfitting and to achieve better generalization \citep{mitchell1980need,mcclelland1992interaction}. Adding inductive bias is supposed to help the system to find desirable solutions without decreasing the performance \citep{battaglia2018relational}.
It can be especially useful for scholarly recommender systems, because they typically face the data sparsity problem: the user ratings are for very few items known, while the number of items compared to the number of users is very large. Indeed, scholarly recommender systems usually have a much higher data sparsity problem than movie recommender systems, for instance, because a large number of scientific articles and venues keeps getting added the database on a daily basis \citep{farber2019microsoft}, and it is not possible for scholars to go through all the articles available. 
\par \textbf{Prevalence:} In scholarly search and recommender systems, there is a scarcity of feedback data on user-item interaction. Vellino examined implicit ratings on Mendeley (which covers research papers) and Netflix (which covers movies) and discovered that Netflix had three orders of magnitude less sparsity than Mendeley \citep{Vellino2013}. Inductive biases are  critical to the ability to classify instances that are not identical to the training instances \citep{mitchell1980need}, for example data on newly added papers, female expert recommendations, etc. \cite{He2017NeuralFiltering} attempt to better generalize the training data using neural networks. To address the data sparsity problem in graph-based recommendation approaches which are mostly used for citation recommender systems, graph embeddings have been proposed in recent works \citep{Tang2015LINE:Embedding,Perozzi2014DeepWalk:Representations}. The latent feature of the nodes capture neighborhood similarity and community membership. Recommender systems that utilize social information for recommendation are also said to mitigate the data sparsity problem \citep{Yang2017SocialTrust}.\\

\item \textbf{Exposure bias} 

\textbf{Definition:} \textit{Users are
exposed to certain relevant items.
However, the unobserved items without interaction do not mean that they are irrelevant or represent negative preference \citep{Chen2020}}.

\textbf{Impact:} Exposure bias is a major issue of implicit feedback datasets. Implicit data does not contain negative feedback: we may know why a user clicked on an item, but we do not know why she did not click on \textcolor{blue}{another}  
item. It could be because she did not find the item relevant, but could also be because she was not exposed to the item, and hence did not ``see'' it. Exposure bias can have the following impact on scholarly recommender systems: (a)~Open-acess (OA) documents and papers from a scholar's own field of study (FOS) are recommended more often \citep{Gupta2021CorrectingRecommendation}. (b)~Exposure bias can exacerbate popularity bias, causing relevant but unpopular items to not be shown \citep{Chen2020}. (c)~Feedback loops (i.e., the recommendation model is trained on interactions of users with an existing recommender system, which might be already biased) also aggravate this bias and diminish the diversity of the recommended results. This could lead to a situation in which users only see a narrow subset of the entire range of recommendations, a phenomenon known as the \textit{filter bubble} \citep{Jiang2019DegenerateSystems}.

\textbf{Prevalence}: For instance, let us consider a recommender system that recommends relevant citations to users to back-up claims \citep{Farber2020CitationDatasets}, showing certain attributes of the paper, such as title, author information, and abstract. In this case, due to the exposure bias, equally relevant \textit{papers from different or rare fields of study} might be less-cited historically (and therefore less recommended) because users have been preferentially exposed to papers in their own field of study, or less often been exposed to papers from a rare field of study. In an observed citation network, a number of relevant papers are not cited because the user was not exposed to those papers \citep{Gupta2021CorrectingRecommendation}. \textit{OA documents} are available to a larger audience, because access is not limited by a “paywall” and can be read by practically anyone that has access to the internet \citep{Harnad2008TheUpdate,Wang2015TheAttention}. This increases their probability of 
being found. \cite{Gupta2021CorrectingRecommendation} research exposure bias in citation recommendation. Upon 
considering
two real-world datasets of citation networks, they use the probability that a node is exposed
to another node as propensity score, and notice that a recommender system trained directly on the observed data underestimates the probability of being cited given exposure for low propensity nodes relative to high propensity nodes. This creates an exposure bias, which can be seen in the lack of diversity in the recommendation results, because papers from the same field of study are recommended almost 95\% of the time, compared to just 82\% of the time when propensity was considered (see Fig.~\ref{exposure_bias}). 
Researchers are mainly interested in the publications in their own disciplines, as \citet{ojasoo1999citation} showed.

\begin{figure}[tb]
    \centering
    \includegraphics[scale = 0.79]{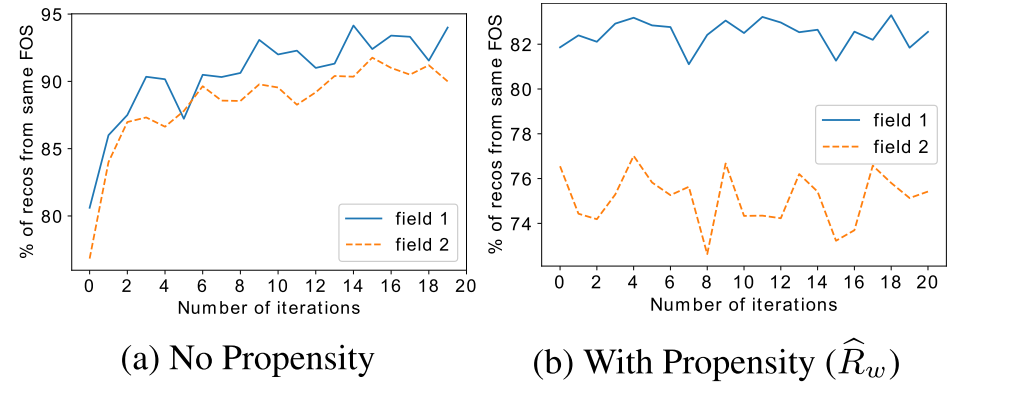}
    \vspace{-0.25cm}
    \caption{Figures from \cite{Gupta2021CorrectingRecommendation} showing the proportions of recommended papers from the field of study (FOS) 1 and 2. For \textit{no propensity} (Fig. 3a), the proportion of recommended papers from the same FOS increases over time for both FOS 1 and 2. This increases exposure bias and decreases the number of recommended papers from a different FOS over time. In contrast, when the models are trained with \textit{propensity} (Fig. 3b), the proportion of recommended papers from a different FOS remains stable over time.}
    \label{exposure_bias}
\end{figure}

The impact 
of exposure bias on popularity in the scholarly field 
can have a similar impact as in other domains. 
For instance, in studying 
 the MovieLens and Last.fm data sets, \cite{Abdollahpouri2020} 
 found that a user is not at all exposed to a vast majority (more than 80\%) of the items.


Exposure bias can affect item diversity. Because of the feedback loops associated with recommender systems, the number of  highly exposed items that are recommended continuously increases over time, and those items that are less exposed despite being relevant get recommended less often with time \citep{Mansoury2020}. On social media, echo chambers describes the lack of diverse perspectives and the favor of the formation of groups of like-minded users, which leads to framing and reinforcing a shared narrative \citep{cinelli2021echo}. Echo chambers can also emerge based on scholarly recommender systems: \cite{Jiang2019DegenerateSystems} show that feedback loops can generate echo-chambers and filter bubbles. They do this by observing the evolution of a user's interest, that is, when a user's interests change extremely over time. Moreover, \cite{Liu2020AData} found that if the data from the previous recommender systems are used in the new recommender without correcting inherent biases, then not only will these biases be carried forward in the new recommender systems model, but they might also be amplified due to the effect of the feedback loop \citep{Chaney2015ARecommendation}.\\

    \item \textbf{Popularity bias}
\par \textbf{Definition:} \textit{The tendency of the recommender system's algorithm to favor a few popular items while under-representing the majority of other items \citep{Abdollahpouri2021User-centeredSystems}. }
\par \textbf{Impact:} Algorithms that prefer popular items are often used to assist people in making decisions from a large number of options (e.g., top-ranked papers in search engine results, highly-cited scientific papers). Recent non-scholarly works \citep{Abdollahpouri2020,SunDebiasingFiltering} show that one of the consequences of popularity bias in search and recommender systems is \textit{disfavoring less popular items}, that is, recommendations are exposed to users purely depending on the degree of popularity. 
This can also create an exposure bias. In the absence of explicit signals or ratings, the systems rely on implicit indications such as popularity and engagement measures. They are simple to use and are frequently used as scalable quality proxies in predictive analytics algorithms. In networks, a bias occurs because all nodes are not equal and the nodes with more links receive more attention. This phenomenon is called ``preferential attachment'' \citep{barabasi1999emergence,barabasi2003scale}. \cite{Sun2007PopularityLibraries} proposed a novel popularity-factor-weighted ranking algorithm that ranks academic papers based on the popularity of the publishing venues. The authors demonstrated that their system outperforms other ranking algorithms by at least 8.5\,\%. The wisdom of the crowd \citep{surowiecki2005wisdom} underpins the utility of such rankings: high-quality options tend to gain early popularity and, as a result, become more likely to be selected since they are more visible. Furthermore, being aware of what is popular can be viewed as a type of social influence; an individual's behavior may be influenced by colleagues or neighbors' choices \citep{Muchnik2013SocialExperiment}. These principles indicate that high-quality material will ``bubble up'' in a system where users have access to popularity or engagement cues (such as ratings, amount of views, and likes), allowing for a 
cost-effective item ranking  \citep{Ciampaglia2018HowQuality}. %
Popularity metrics 
have not only been adopted 
to popular social media and e-commerce platforms with the general population as users, 
but they are also used on social media platforms like paperswithcode\footnote{\url{https://paperswithcode.com/}} for researchers to highlight popular and trending items. Many recommender systems disregard unpopular or recently released items with minimal ratings, focusing instead on those with enough ratings to be useful in recommendation algorithms \citep{Park2008TheIt}. It is important to tackle popularity bias due to the following reasons:
\begin{enumerate}
    \item Popularity bias reduces the extent of \textit{personalization} and harms \textit{serendipity} \citep{Abdollahpouri2017ControllingRecommendation}. In addition, the \textit{diversity} of the recommendation list could be decreased \citep{Abdollahpouri2019ManagingRe-Ranking}. Different users have distinct preferences. Only recommending popular papers, authors, citations, etc. might not always be beneficial especially for those users that study or wish to research a rare or unexplored subject/field of study. There would also be no surprise element for the user, as she keeps seeing items that she has already seen before. Thus, the recommendation is not based on the actual content, but based on effective rules of thumbs. 
    \item \textit{Unfairness} of recommendation results also increases \citep{Chen2020}. Even if they are good matches, regularly recommending popular items diminishes the exposure of other items, which is unequal with respect to the less popular items. 
    \item Popularity bias will boost the exposure opportunities of other popular items \citep{Abdollahpouri2020}, making them even more popular. As a result, the training data also becomes unbalanced, elevating the so-called \textit{Matthew effect}.

\end{enumerate}

    \begin{figure} [tb]
    \centering
    \includegraphics[scale = 0.66]{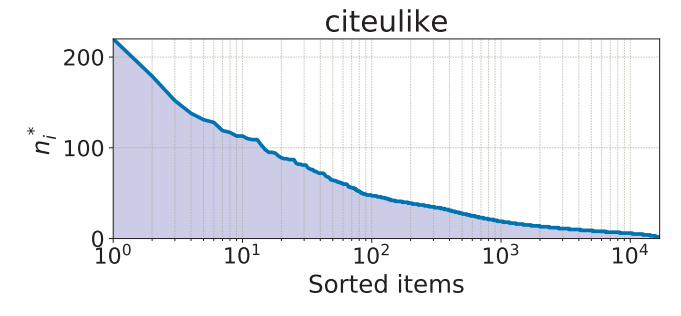}
    \caption{Figure from \citep{Yang2018UnbiasedFeedback}. The distribution of $n_i^*$ (the observed number of interactions with item $i$) in the citeulike dataset. The items are presented in descending order of $n_i^*$. The horizontal axis is log scaled for better visualization. The $n_i^*$ distribution is skewed and the user interactions are significantly biased.}
    \label{Popularity1}
    \end{figure}
\par \textbf{Prevalence:} Quite recently in a paper by \cite{Zhu2021Popularity-OpportunityFiltering}, popularity in terms of equal opportunity instead of the conventional popularity bias was considered. According to this paper, given that a user likes both a popular item $i$ as well as a less popular item $j$, $i$ is observed to rank higher than $j$ even though the user likes both items. Due to position bias exhibited by the user, item $i$ is more likely to be observed and clicked than item $j$, resulting in bias favoring popular items.\\
\cite{Yang2018UnbiasedFeedback} investigated two types of bias in terms of popularity in the CiteULike dataset, which contains user ratings for scientific publications: (a) \textbf{interaction bias} (i.e., the tendency for users to connect with popular items more frequently), and (b) \textbf{presentation bias} (i.e., the tendency for recommenders to unfairly present more popular items than long tail ones). Though this differentiation was used to highlight the prevalence of popularity bias in scientific articles in the CiteULike dataset, it can also 
be made for all kinds of scholarly items, such as venues, citations, and experts.  \textit{Interaction bias} can be seen clearly in Figure \ref{Popularity1}. The $n_i^*$ distribution is highly skewed because the horizontal axis is log scaled: Less than 50 user interactions occur for 99 percent of the goods. However this kind of bias can be rather well understood from a user perspective as it largely depends on users' behavior and her decision on which items she chooses to interact with. This can also be seen as a result of conformity as users tend to interact with popular items, because these items are most favored by majority of other users, and therefore they choose to \textit{follow the crowd}. For this very reason, we decided to include interaction bias as a type of conformity bias, which we have explained in detail previously (see Sec.~\ref{Conformation}).

\par \cite{Yang2018UnbiasedFeedback} calculated the average number of times that an item with the observed popularity $n^* \in [1, max(n_i^* )]$ was recommended, denoted by f($n^*$), in order to account for the presentation bias. A biased recommender may provide a $f(n^*)$ that is linearly or exponentially rising, whereas an unbiased system should anticipate one that is basically flat with a small slope (see Fig.~\ref{Popularity2}). In the scholarly field, popular items are not only limited to highly-cited research papers, but can also include famous authors, conference venues, academic events, etc. \\ 
    \begin{figure} [tb]
    \centering
    \includegraphics[scale = 0.8]{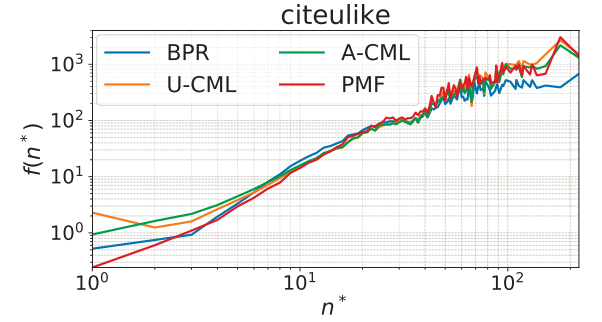}
    \caption{Figure from \citep{Yang2018UnbiasedFeedback}. Empirically estimated f($n^*$) on the CiteULike dataset and four recommendation algorithms. $f(n^*)$ denotes the average number of times that an item with observed popularity $n^*$ was recommended. Both axes are log scaled. Therefore, exponential growth is linear in the figure. This shows significant presentation bias.}
    \label{Popularity2}
    \end{figure}

\item \textbf{Unfairness}
\par \textbf{Definition:} \textit{The system systematically and unfairly discriminates against certain individuals or groups of individuals in favor of others \citep{Chen2020}}. 
\par \textbf{Impact:} Unfairness is a serious and ongoing issue in academia where certain groups are discriminated based on gender \citep{Holman2018TheRepresented}, race \citep{Fang2000RacialMedicine}, socio-economic status \citep{Petersen2014ReputationCareers}, and university prestige \citep{Way2019ProductivityEnvironment} among many others. Such inequalities have been found in grant funding \citep{Lee2015GenderNetherlands}, credit for collaborative work \citep{Sarsons2017RecognitionAcademia}, hiring and promotions \citep{Nielsen2016LimitsProcesses}, peer-review \citep{Shah2021AnReview}, authorship \citep{West2013TheAuthorship} \citep{Huang2020HistoricalDisciplines}, and citations \citep{Caplar2017QuantitativeCounts}. The difficulty of fairness has prevented recommender systems from becoming more pervasive in our culture. Particularly, distinct user groups are typically underrepresented in data depending on characteristics like gender, affiliation, or degree of education. The recommender systems models are quite likely to learn these overrepresented or underrepresented groups when training on such unbalanced data, reinforce them in the ranked results, and possibly lead to systematic discrimination and decreased visibility for disadvantaged groups. Moreover, due to the effects of using historical data in supervised learning settings (see also the feedback loop for recommender systems, i.e., using the information of user behavior with recommender systems to create recommender system models), these results have the potential to (unknowingly) influence user thinking and behavior, and also make users part of the \textit{filter bubble}, again re-enforcing these beliefs. These biases amplify over time due to the \textit{Matthew effect}.
    
\par \textbf{Prevalence:} \cite{Andersen2018GoogleDisciplines} observed an indirect \textbf{gender bias} in the Web of Science's selection criteria for its citation resources. \cite{Mohammad2020GenderCitations} observed gender gaps in Natural Language Processing (NLP) both in authorship as well as citations. Among all the NLP papers published from 1965 to 2000 in the ACL Anthology,\footnote{\url{https://aclanthology.org/}} they found that only about 30\% were written by women. They also used citation counts extracted from Google Scholar to show that, on average, male-first authors are cited markedly more than female-first authors, even when the experience and area of work are considered. Recommendations in educational and career choices are another important motivating application for fair recommender systems. 

Students’ academic choices can have significant impacts on their future careers and lives. \cite{Polyzou2020ModelsEducation} found a course recommender system to give biased results to a student because of pre-existing biases and socio-technical issues present in input data. For instance, a system 
might rarely advise female computer science students to take intensive coding programs. If patterns from the last 50 years continue, parity between the number of male and female authors will not be attained in this century, according to Wang et al.'s detailed examination of gender trends in the computer science literature. They came to the conclusion that there is a persistent gender gap in the literature on computer science that might not be closed without deliberate action \citep{Wang2021GenderAuthorship}. Google Scholar is of the most popular academic search engines which uses citation counts as the highest weighted factor for ranking publications. Therefore, highly cited articles are found significantly more often in higher positions than articles that have been cited less often \citep{Beel2009AcademicAlgorithm}. According to Parker et al., a small group of scholars receive a disproportionately significant amount of citations in the scientific community. They looked at the social characteristics of these highly cited scientists and discovered that the majority of them are male, nearly entirely based in North America and Western Europe \citep{Parker2010CharacterizingEcology}. Most search and recommender systems for scholars therefore show unfairness against less-cited authors, gender, and race, among many others in the recommended results.
\par Bias in \textbf{expert recommendation} within academia has been researched relatively fairly by many scholars. Expert recommender systems have been used for the following (see \citep{Salman2020IncorporatingRecommendation}): (1) to hire and recruit professors and researchers in academic positions, (2) to recommend experts to evaluate patents, (3) to identify reviewers for scientific conferences, and (4) to assemble a conference program committee. One study by the Nature Magazine \citep{Lerback2017JournalsReferee} shows that only 20\% of the peer-reviewers are women and are thus largely under-represented. Another study \citep{Yin2011FindingNetworks} shows that women and authors from developing countries were under-represented as editors and as peer-reviewers. The National Science Foundation (NSF) developed an automated reviewer selection system that considers different demographic features while selecting peer reviewers \citep{Hettich2006MiningReviewers} to tackle this problem. Bias in a peer review process can also be seen in the geo-location of the reviewer. For example, a study in \citep{Publons2018GlobalReview} shows that the US dominated the peer review process by 32.9\% while its publications represent only 25.4\% of all publications. %
\par Unfairness in \textbf{language} of scientific publication can also be observed. Journals not published in English tend to have a lower impact factor \citep{Vanclay2009Impact}. Matías-Guiu and García-Ramos reason that this may not be due to the language itself, but to the fact that they are not included in authorship networks, since most English speaking authors do not read them, and hence, do not cite them either \citep{Matias-Guiu2011EditorialPublications}. \cite{May1997TheNations} also found serious English language bias in the Institute for Scientific Information (ISI) database, as most non-English journals are not covered by the Science Citation Index (SCI). 
Cross-lingual citations make up less than two percent of all citations \citep{cross2021}. 
In the field of recommender systems, \cite{Torres2004EnhancingTechLens} observed that a recommender in a user's native language was greatly preferred to one in an alternative language, even if the items themselves recommended were in the alternative language (e.g., a Portuguese-based recommender recommending scientific papers in English).

Another important instance of unfairness in the scholarly field is the phenomenon of \textbf{citation bias}. Citations are key elements in the evolution of scientific knowledge and of the scientific discourse. They enable particular research findings to survive over time and to develop into academic consensus. Given the large body of scientific literature, 
it is often unfeasible to cite all published articles on a specific topic, particularly in case of page limitations given by journals and conference proceedings, although the San Francisco Declaration on Research Assessment (DORA; \url{http://www.ascb.org/dora/}) encourages publishers to not insist in such constraints). As a result, some selection of citations needs to take place. If this selection is influenced by the actual results of the research presented in the article, then citation bias occurs \citep{song2010dissemination}. \cite{Duyx2017ScientificMeta-analysis} conducted a systematic review and meta-analysis on the citation behavior of about 52 studies from the Web of Science Core Collection and Medline.\footnote{\url{https://www.medline.com/}} They found that positive articles, that is, articles that support an author's belief/claim, are cited about twice as often as negative ones, i.e., articles that are critical to an author's belief/claim. \cite{Greenberg2009HowNetwork} conducted a similar analysis and found that positive articles received 94\% of the total citations. %

\end{enumerate}

\textbf{Summary.} Overall, we can highlight the following aspects concerning biases in scholarly recommender systems:
\begin{enumerate}
 \item All biases concerning recommender systems in general are also relevant to scholarly recommender systems.
 \item In particular, \textit{unfairness} in the form of gender bias has been addressed 
due to unbalanced gender distribution in academia \citep{sapiezynski2017academic,islam2021debiasing}. 
 \item While scholarly 
recommender systems aim to provide diversity in terms of content, 
they might also be responsible for ``echo chambers'' in science \citep{Weste1912444117}
due to the \textit{exposure bias} in underlying data. This is frequently observed in citation networks \citep{Gupta2021CorrectingRecommendation}. 
 \item The impacts of the \textit{Matthew effect} cannot be ignored: most search and recommender systems display their results in an order influenced by citation counts and related criteria, and frequently cited papers are cited disproportionately more often as they increase in popularity.
\end{enumerate} %

\section{Bias Mitigation in Scholarly Recommender Systems} %
\label{sec:methods-mitigate}

\subsection{Scholarly Bias Mitigation Methods}

According to \citet{DAlessandro2017}, bias mitigation methods can be grouped into the following three categories. They align with the steps of recommender systems' lifecycles, mitigating bias in the data collection, training, and recommendation steps (see Sec.~\ref{Label5}):
\begin{enumerate}
    \item \textbf{Mitigation during preprocessing}: This type of mitigation approach is deployed before the creation of a recommender system's model. Using these mitigation techniques, the dataset is transformed with the aim of removing the underlying bias from the data before modeling. However, these methods are usually hard to implement and may be ineffective, as they cannot remove the bias present in the machine learning model itself. These techniques are mostly used to decrease the level of discrimination and unfairness in the data \citep{Calmon2017OptimizedPrevention}.
    \item \textbf{Mitigation during processing}: This mitigation approach employs techniques that can be considered modifications of the traditional learning algorithms to address bias during the model training phase (e.g., utilizing bias-indicating metrics, such as propensity scores).
    \item \textbf{Mitigation during postprocessing}: This mitigation approach uses techniques in which bias processing is performed after model training. These mitigation methods are noteworthy because the user's perception of and interaction with the results are essential. 
\end{enumerate}

\begin{figure}[tb]
   \centering
   \includegraphics[width=0.8\textwidth]{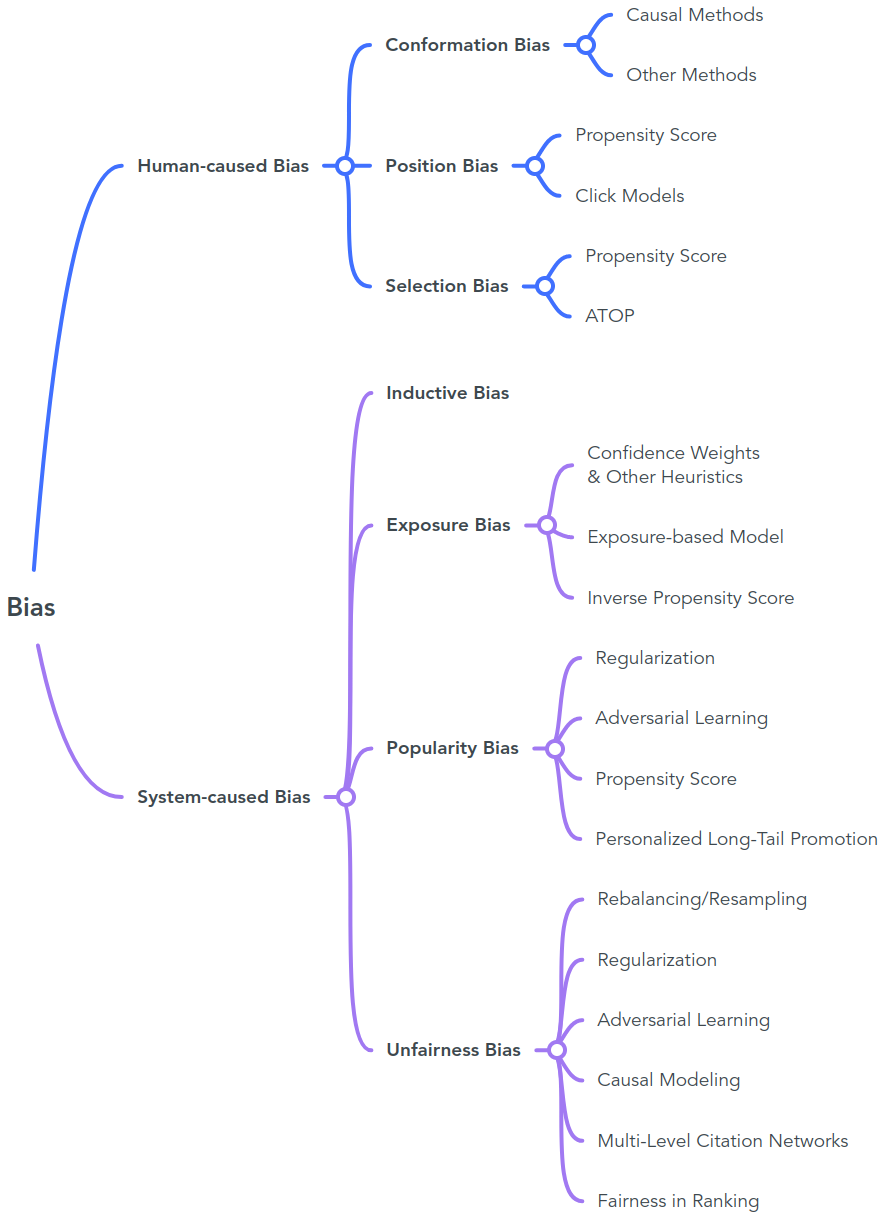} 
   \caption{Overview of identified bias mitigation methods.}
   \label{fig:Mitigation}
\end{figure}

Fig.~\ref{fig:Mitigation} and Tables \ref{mitigation_publication} and \ref{mitigation_publication-part2} provide an overview of the mitigation methods mentioned in our paper collection. 
Overall, we see that concrete bias mitigation methods have been published with respect to all biases types outlined in Sec.~\ref{sec:biases-rec}. As dealing with bias during model training requires particular methods (e.g., neural network architectures), a large portion of publications is dedicated to within-model bias.

\renewcommand{\arraystretch}{1.4}
\begin{sidewaystable}[ph!] %
\vspace{12cm}
\begin{tabular}{|L{3.2cm}|L{1.4cm}|L{2.2cm}|L{10.9cm}|}
\hline
\textbf{Bias} & \textbf{Stage in Rec.Sys.} & \textbf{Mitigation Method}  & \textbf{References} \\ 
\hline
\hline 
\multirow{2}{*}{\shortstack[l]{\\\textbf{Conformation bias}\\ (Unreliable feedback \\data)}} & In  & Causal embedding & \cite{Zheng2020DisentanglingEmbedding} \\ 
\cline{3-4} 
 &  & Other methods & \cite{Wang2017LearningRecommendation,Chaney2015ARecommendation,Tang2012MTrust:World,Ma2009LearningEnsemble}  \\ 
 \hline
\multirow{2}{*}{\shortstack[l]{\\\textbf{Position bias} \\(unreliable click data)}}    & in & Propensity score & \cite{Agarwal2019ALearning-to-rank,Fang2019InterventionEstimation,Agarwal2019EstimatingInter-ventions,Ai2018UnbiasedEstimation,Joachims2017UnbiasedFeedback,Hu2017ARank,Swaminathan2015BatchMinimization,Raman2013LearningUsers,Chapelle2012Large-scaleEvaluation} \\ 
\cline{2-4} 
  & other & Click models    & \cite{Vardasbi2020CascadeRank,Borisov2016ASearch,Shen2012PersonalizedFiltering,Xu2012IncorporatingModels,Chapelle2009ARanking,Dupret2008AObservations,Craswell2008AnModels} \\ 
  \hline
\multirow{2}{*}{\shortstack[l]{\\ \textbf{Selection bias} \\(unreliable positive \\feedback data)}} & In & Propensity score & \cite{Saito2020,Schnabel2016RecommendationsEvaluation} \\ 
\cline{2-4} 
  & post & ATOP  & \cite{Lim2015Top-NFeedback,Steck2010TrainingRandom}  \\ 
  \hline
\end{tabular}
\caption{Mitigation methods for human-caused biases.} %
\label{mitigation_publication}
\end{sidewaystable}

\begin{sidewaystable}[ph!]
\begin{small}
\vspace{12cm}
\begin{tabular}{|L{2.5cm}|L{1.2cm}|L{2.55cm}|L{11.5cm}@{\hskip3pt}|}
\hline
\textbf{Bias} & \textbf{Stage in Rec.Sys.} & \textbf{Mitigation Method}  & \textbf{References} \\ 
\hline
\hline 
\multirow{3}{*}{\shortstack[l]{\\\textbf{Exposure bias} \\(unreliable negative \\feedback data)}} &
in & 
Confidence weights \& other heuristics &  \cite{Saito2020,Chen2018MissingRecommendation,Sidana2018LearningPANDOR,Lian2017MutualRecommendation,He2016FastFeedback,LiImprovingInformation,Pan2009MindFiltering,Hu2008CollaborativeDatasets,Pan2008One-classFiltering}                               \\ 
\cline{3-4} 
 & & Sampling & \cite{Liu2019DistributedRecommendation,Zhang2018Task-guidedInference,Karvelis2018TopicDoc2Vec,Caselles-Dupre2018Word2vecMatter,Rendle2012BPR:Feedback}  \\
\cline{3-4} 
  & & Exposure-based model & \cite{Pradhan2020CNAVER:System,Liang2016} \\
\cline{2-4} 
 & post & Propensity score & \cite{Gupta2021CorrectingRecommendation,Yang2018UnbiasedFeedback} \\
\hline
\multirow{3}{*}{\shortstack[l]{\\\textbf{Popularity bias} \\(popular items \\get recommended \\more often)}} & in & Regularization  & \cite{Abdollahpouri2021User-centeredSystems,ChenESAM:Performance,Abdollahpouri2017ControllingRecommendation} \\ 
\cline{2-4} 
  & in & Adversarial learning & \cite{Krishnan2018AnFiltering-1.5pt} \\
  \cline{2-4} 
 & in & Propensity score  & \cite{Yang2018UnbiasedFeedback} \\ 
 \cline{2-4} 
  & post & Personalized long-tail promotion & \cite{Abdollahpouri2019ManagingRe-Ranking} \\ 
  \hline
\multirow{3}{*}{\shortstack[l]{\\\textbf{Unfairness} \\(unfairness to \\certain groups)}} & pre & Re-balancing/ Re-sampling & \cite{CemGeyik2019Fairness-AwareSearch,Asudeh2019DesigningSchemes,Biega2018EquityRankings,Pedreshi2008Discrimination-awareMining} \\ 
\cline{2-4} 
 & in & Regularization & \cite{Morik2020ControllingLearning-to-Rank,Beutel2019FairnessComparisons,Burke2018BalancedRecommendation,Kamishima2017ConsiderationsTask,Yao2017BeyondFilteringb,Abdollahpouri2017ControllingRecommendation,Kamishima2016Model-BasedRecommendation,Zemel2013LearningRepresentations,Kamishima2013EfficiencyRecommendation}  \\ 
 \cline{2-4} 
 & in & Adversarial Learning & \cite{Bose2019CompositionalEmbeddings,Edwards2015CensoringAdversary}       \\ 
 \cline{2-4} 
 & in  & Causal Modeling & \cite{Zhang2018FairnessFormula,Nabi2018FairOutcomes,Wu2018OnGraph,Kusner2017CounterfactualFairness} \\ %
 \cline{2-4} 
 & in & Multi-level citation networks          & \cite{Son2018AcademicNetworks}      \\
 \cline{2-4} 
 & post & Fairness in ranking & \cite{Biega2020OverviewTrack,Biega2018EquityRankings,Yang2017MeasuringOutputs,Zehlike2017FAIR:Algorithm} \\ 
 \hline
\end{tabular}
\end{small}
\vspace{-0.1cm}
\caption{Mitigation methods for system-caused biases.} %
\label{mitigation_publication-part2}
\end{sidewaystable}

\subsection{Debiasing Framework} %
\label{Extra_main}

Notably, most bias mitigation approaches in recommender systems only focus on one bias dimension. Only one paper \citep{Polonioli2020TheSystems} addresses three types of bias while focusing on the ethics of recommender systems. However, in reality, not just one but many types of bias affect recommender systems' performance 
simultaneously. 
Thus, in this paper, to mitigate the mixture of biases found during the steps of recommender systems' lifecycles, 
we provide 
a universal debiasing framework for scholarly recommendation. 

Our debiasing framework is 
designed for the \textit{scholarly domain}. 
As outlined in Section 1, in contrast to non-scholarly 
recommender systems (e.g., movie recommender systems), scholarly 
recommender systems have various unique characteristics. 
For instance, the content (full-text) of papers and the topology of networks (e.g., citation networks) are subject to scholarly biases. In addition, researchers, institutions, and papers are embedded into scientific communities and societal behavior. %
Technically, scholarly recommender systems typically need to deal with noise and the lack of negative feedback.
Overall, our debiasing framework focuses on identifying and finding solutions pertaining to 
the mentioned issues that are frequently observed in scholarly AI systems. %

Our debiasing framework consists of three steps: %

\begin{enumerate}
    \item \textbf{Detect biases:} To debias a recommender system, it is first important to detect the kinds of biases that occur (and that can be measured). 
    This can vary depending on the use case.\\
    To determine what biases a system is dealing with, it is important to know the number of false negatives 
    and false positives 
    from the feedback data. These numbers can be determined via a simple confusion matrix or standard evaluation metrics, such as precision, recall, etc. User click-data can also be helpful.  
    We can list the following 
    characteristics of each bias type as rules of thumb to determine the type of bias in the data: \\ %
    (i) Exposure bias: if the number of false negatives 
    is too high;\\
    (ii) Position bias: if the click data are imbalanced;\\
    (iii) Selection bias: if the number of 
    false positives is too high;\\
    (iv) Conformation bias: if click data are similar within a community.\\
   This list is only meant to serve as a brief orientation. To detect and mitigate biases in recommendation results, popularity and fairness metrics such as statistical parity (or group fairness; \cite{Biega2018EquityRankings}), precision, and recall \citep{Wang2011CollaborativeArticles} need to be determined as well. For a detailed explanation on how to figure out the types of bias a recommender system faces, 
   please refer to our descriptions in Sec.~\ref{sec:biases-rec}.

    \item \textbf{Selection of mitigation method(s):} Once the bias(es) has been found, one can refer to Tables~\ref{mitigation_publication} and \ref{mitigation_publication-part2}, as well as to a detailed description in Sec.~\ref{sec:methods-mitigate}, to select suitable mitigation methods. 
    Developers can choose between methods that are performed before, during, or after data processing. This helps mitigate bias in the data collection, training, and recommendation steps (see Sec.~\ref{Label5}). 
    \item \textbf{Model assessment and evaluation:} After applying the appropriate mitigation method, a recommender systems model can be evaluated against baselines to evaluate whether its recommendation results are debiased and its performance changed. %
\end{enumerate}

\textbf{Example Use Case.} To illustrate the need for and usage of a debiasing framework for scholarly recommender systems, let us consider the following scenario. 
Data scientists at the digital library ACM\footnote{\url{https://dl.acm.org/}} might notice a decline in the click-through rate (CTR) of their recommender system. They 
perform an offline evaluation and find that their model fails to beat simple baselines, concluding that their recommendation algorithm is still biased despite using general techniques like regularization 
during model training. 
They realize that they need a comprehensive framework to debias their recommender system with respect to several biases. By analyzing the click data, they find that the click data is highly imbalanced, but that it forms clusters. Thus, both \textit{position bias} and \textit{conformation bias} need to be addressed. 
Our framework (see Table 1 and 2) then reveals that \textit{position bias} can be reduced by introducing a propensity score (e.g., \cite{Agarwal2019ALearning-to-rank,Fang2019InterventionEstimation,Agarwal2019EstimatingInter-ventions}) or by using click models (e.g., \cite{Vardasbi2020CascadeRank}), while \textit{conformation bias} can be reduced in various ways, such as by means of causal embeddings \citep{Zheng2020DisentanglingEmbedding} as one of the most recent approaches. %

\textbf{Example Use Case 2.}
Finding suitable researchers to review papers is often not an easy task for editors. Several academic publishers (e.g., Elsevier) offer online recommendation systems to find suitable reviewers for a given paper. However, since the editor's selection of reviewers can be slightly biased (e.g., selecting people who already have a high h-index or have already reviewed many papers in the journal, i.e., popular people), it makes sense to determine suitable debiasing methods. The present case indicates that the \textit{popularity bias} should be reduced. In the event that the publisher does not develop the recommendation system itself, but only has a license and obtains the recommendations via an API, a mitigation method during post-processing should be looked into. In this case, our framework (Table 2, Popularity Bias, Stage ``post'') suggests that the personalized long-tail promotion method of \cite{Abdollahpouri2019ManagingRe-Ranking} can be used by the publisher to re-rank the results before displaying them to the users.

\section{Future Directions and Open Challenges}
\label{Label27}

In this section, we discuss the open challenges that scholars might face while building bias-aware recommender systems for academia, and hope that this will stimulate 
future research concerning scholarly biases and their mitigation in scholarly recommender systems. 
We do this by transferring the challenges mentioned by \cite{Chen2020} 
to the scholarly domain and %
by introducing additional aspects, such as using scholarly knowledge graphs to provide explainable scholarly recommender systems. 

\textbf{Nontrivial Statistical Techniques.} %
Inverse propensity scoring (IPS) is a technique widely used to provide an unbiased estimate of the ranking metric of interest and to re-weight clicked items 
\citep{Wang0WW21}. IPS-based methods are the most frequently used bias mitigation method in recommender systems in general \citep{Chen2020}. 
However, these methods are only effective if the propensities are correctly estimated. An important 
question here is how propensity scores can be estimated accurately. 
Although some proposed methods provide 
accurate estimation given a simple bias, such as position bias \citep{Swaminathan2015BatchMinimization,Chapelle2012Large-scaleEvaluation}, it can be difficult to estimate propensity in the presence of complicated biases, such as exposure and selection biases \citep{Lee2021DualFeedback}. Unlike position bias, these do not simply rely on the position of an item, but 
on many other factors such as item popularity, availability (open access vs non-open access items), and feedback loops. 
Most publications on bias in our paper collection do not address this issue. Therefore, we support \citet{Chen2020}'s argument 
that further research needs to be performed in this area, particularly in the scholarly field with its typical sparse data (e.g., limited amount of click data).

\textbf{Addressing Biases Simultaneously. } 
We have noticed that the various methods proposed to mitigate bias in %
recommender systems usually address only one or at most two biases, while 
in reality multiple types of bias might occur simultaneously. For instance, to decide whether to download a paper, a researcher might be influenced by the people in her scientific community due to conformation and popularity biases. 
We found few papers that address multiple biases simultaneously. \cite{Chen2021AutoDebias:Recommendation}, for instance, used meta-learning techniques to learn the debiasing parameters from training data. \cite{Jin2020} provided a transfer learning approach that mitigates one or multiple biases in downstream classifiers by transfer learning from an upstream model. 
To tackle the issue of multiple bias types systematically, two solutions seem to be viable: (1)~a general debiasing framework (which we provide in this article) to identify all necessary bias mitigation methods that need to be applied one by one; or (2)~a mitigation method that can tackle multiple biases simultaneously---and ideally, model and recognize dependencies between biases. 

\textbf{Addressing Bias in Practice. } %
While debiasing methods have been proposed in research, 
methodologies and guidelines 
for practitioners to address bias in real use cases (e.g., software developers of recommender systems and data scientists) is largely missing. %
While we provide a debiasing framework in this article, we see a need to provide more software libraries to identify and mitigate biases. %

\textbf{Addressing Bias in Evaluation.} %
Evaluating bias is a tricky endeavor, as detecting and mitigating specific biases depends greatly on the available data. 
Most evaluation methods either require accurate propensity scores or unbiased click data. Both can be difficult to obtain in implicit feedback data. Uniform data is sampled with equally-sized sets from each class. Although uniform data provide unbiased information, their small scale and price are insufficient and expensive to evaluate recommender systems, mainly due to high variance found in the data. Moreover, due to the increasing presence of popularity bias and unfairness in the data (see Sec.~\ref{sec:biases-rec}), many works propose different kinds of evaluation methods, which can be inconsistent across different kinds of results. Therefore, more research 
into formulating 
new metrics for evaluating 
recommender systems is needed, with or without biased click data. 

\textbf{Deeper Understanding and Modeling of User Behavior. }
Recommender system biases are %
caused by human factors, as humans use the recommender systems and typically provide historical data for model training. 
Thus, we argue that it is worthwhile to model and understand the users themselves (e.g., researchers grouped by experience and scientific discipline) in their daily working environments. For instance, we can assume that paper and citation recommendation systems will be well received in the future. However, these systems are assumed to be 
prone 
to biases \citep{Farber2020CitationDatasets}. %

Specifically, citation bias has been analyzed in two main contexts in the literature: to explain the scholars' self-citation behavior, and 
to show that scholars cite papers but disproportionally criticize papers or specific claims less often. 
While the first phenomenon has been addressed 
in a few journal reports \citep{VanNoorden2019HundredsDatabase,Aksnes2003ASelf-citation}, little literature can be found to tackle the latter issue, which deals with the context in which a scholar cites. To solve this problem, it is important to understand scholars' citation behavior. 

\textbf{Explainable Recommender Systems.}  
Making recommender systems more explainable has been considered key in recent years. This is particularly important in the scholarly field, one which requires a high level of trust. With respect to biases, explanations of recommendations can help users make informed decisions---facilitating human agency---and can reduce certain biases via human-computer interaction. Existing frameworks and ethics guidelines for trustworthy AI \citep{JobinIV19}, such as the one provided by the European Union \citep{trustworthyAI}, can be implemented here.

In addition, scholarly knowledge graphs are worth 
mention 
as a way of modeling academic knowledge explicitly, paving the way for scholarly knowledge graph-based recommender systems \citep{Ayala-Gomez2018GlobalGraphs}. For instance, the Microsoft Academic Knowledge Graph (MAKG)\footnote{Microsoft Academic website and underlying APIs was retired on Dec. 31, 2021. \url{https://www.microsoft.com/en-us/research/project/academic/articles/microsoft-academic-to-expand-horizons-with-community-driven-approach/}} \citep{farber2019microsoft} contains eight billion triples about publications and associated entities, such as authors, venues, and affiliations. Wikidata (\url{https://wikidata.org}), OpenCitations \citep{DBLP:journals/qss/PeroniS20}, the Open Research Knowledge Graph (ORKG) \citep{ORKG}, and AIDA \citep{AIDAqss}, among others, are further noteworthy knowledge graphs. %

\section{Conclusion}

In this paper, we provided readers with an overview of biases present in scholarly recommender systems. In particular, we outlined the impact and prevalence of each type of bias. We then investigated which of the published bias mitigation methods are particularly relevant in the context of scholarly recommender systems. 
We noticed that scholarly biases have been rather underexplored so far, particularly in the context of scholarly recommender systems. 
It became clear that we easily run the risk of 
leaving biases undetected in academia, potentially disadvantaging millions of researchers. 
Therefore, we proposed a simple-kept framework that readers can follow to help them choose the right mitigation method 
for their scholarly recommender systems. 
Last but not least, we made suggestions for future 
research in this area.

\section*{Conflict of Interest}
The authors declare that they have no conflict of interest.

\bibliographystyle{spbasic_mod}      %
\bibliography{extracted}   %

\end{document}